# BitUP: Efficient Bitmap Data Storage Solution For User Profile

Derong Tang, *Tecent*   Hank Wang*, Tencent*

## Abstract

User profile is widely used in the internet consumer industry, it can be used in recommendation systems for better user experience [1], or improving Ads system with better conversion rate [2]. Most internet situation we must met large scale data set, thus retrieve efficient and store with less space became a challenge, how to handle trillions rows of data is very common in our business scene, so we proposed a novel solution called **BitUP,** involved the new distributed bitmap structure to efficient store profile label data, and can querying with better performance, the fundamental structure is bitmap, that's why our method is efficient and less storage overhead.

Our design can also scale linearly, to demonstrate the scalability of the proposed solution we show retrieval efficiency in various industry data, which scales across from terabytes to petabytes.

## 1. Introduction

User Profile is critical in all customer-based systems. How to model user profile data or store it was quite critical to support business development. Along with the growth of business, the data could scale up to billions. We use bitmap to store profile data for better performance and lower cost. There are three challenges for consumer-based profile data:

- The data came from heterogeneous source like various device, different warehouse or db, moreover from different country

- Super large scale, most commonly need distributed store system, so how to design the base data structure is critical

- Many aspects need to depict user behavior for better business value, so if you use a wide table to store, the column size may exceed 10 thousands, it would take much effort to get the final widetable.

Our method uses a distributed bitmap structure to accomplish all of the above challenges, running smoothly in the real industry data, and avoiding the intensive workload in the wide table merge process, moreover it also can scale up easily.

### 1.1. Contributions

Since its previous introduction, the **BitUP** system has proliferated across various industry business scenes, solved scale problems and widely used in recommendation and ads systems for user experience. It provide a robust, versatile, large scale, fault tolerance solution; Our system contains four advantages:

**Avoid Join Process:** Typical solution for combine heterogeneous source data is by join, then you get a wide table with different data together, but there are several drawbacks, at the first place you need additional resource or computation to join the data together, when the source or data scales up, the cost is pretty high and not acceptable. Our approach just uses the source table to construct the store bitmap, taking advantage of bitmap structure, which could later merge data by **AND, OR, XOR** operations in memory.

**Efficient Bitmap Data Structure:** The largest data scales reach to petabytes, which seriously affect the performance to retrieve, therefore we proposed a distributed bitmap method to increase parallel level in query stage, and introduced tablet concept to better abstract the storage layer in OLAP table engine.

**Consecutive ID Generator Across Days:** From practical experience, the industry profile data involved various id systems, some maybe numbers like user id, most of others were device strings that came from hash algorithms such as oaid, imei, or caid. However the bitmap only receives number representation in its bit store. Therefore we presented a novel ID Generator to map string id to number, after that it could be stored into bitmap. Another ability we achieved, to prevent the id change across different days, the same origin id must be the same number id in any period that can assure the follow-up flow can run consistently, locate the right person and use the correct user profile.

**Data Quality Assurance:** The data quality is quite critical in a data system, we introduce a whole bunch of metrics to measure the quality of user profile.

Notably, The novelty of this work is the newly presented distributed bitmap structure that can solve industrial scale problem, and picking the right related solution and databases implement it, with the novelty structure we can reduce large amount of effort previous need such as wide table join, and it's a leading level user profile platform in industry, also used in Tencent.

**1.2. Related Work**

Related works have been implemented in Alibaba, ByteDance, Kuaishou and so on. [1, 2, 3, 4, 5] all shows the user profile system is a critical part in recommendation or ads work flow, the approach of [7] is quite outdated, and can not use the modern ability like vector engine, new compression method, and bitmap support.

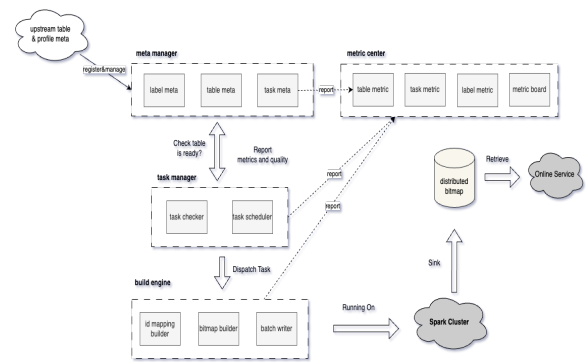

Figure 1: The overall architecture of **BitUP** system

## 2. Architecture

The user profile system, always called as a data management platform. This whole system is really complicated. It manages the flow of data production, the quality of ingest data, and also the pipeline to online service, furthermore accelerate this data flow from raw data to online business, help data produce business value with little effort, easy use for users.

The whole system contains different modules as shown in Figure 1. The upstream source table came from various heterogeneous systems. The profile system needs to contain the metadata of each single table, the column, the upstream task yield this table, and the owners. All of the above will register in our meta manager, this module will manage and save all the metadata in our system, providing important information for follow-up workflow. Meta managers also collect quality metrics from downstream tasks, illustrate the whole lifecycle metrics of any profile label or table. It's quite important to answer several questions: **where is my label, and what's the state, producing, writing or already serving? If it works well in business?**

Task manager is a route center to check which table is ready for build, and then dispatch a specified builder task of this table such as id mapping or bitmap. It retrieves table meta from the meta manager and checks the upstream task state in a fixed duration.

Build engine is a core module in the current system, it encompass three parts with different tasks: the id mapping builder will convert the string id to number, the bitmap builder will construct a distributed bitmap from upstream table with number id converted, and the batch writer will sink to various kinds of OLAP stores with efficient batch mode. Furthermore the whole engine runs on a **Spark** [8] cluster because the above building task is implemented by Spark.

### 2.1. meta manager

In practical industry profile data, always contains different source table at scale, if you handle it case by case, you can hardly manage it smoothly, even with much effort, therefore we introduce meta manager to register and manage the complex relation between upstream tables and final profile label, register the information such as the mapping column to label, the table task which following task depends on, and the **UDF** writer relied on.

We abstract this workflow by several relation model, represent the source metadata, the label meta, after that the whole system could query lineage across task, table and label by relation model, as a consequence, the system could indicate when parent task has been done, the empty ratio or feature length of every label, moreover if the profile satisfy SLA required by downstream business. All the pipeline could be automated with little human effort.

| user id | age | gender |
| --- | --- | --- |
| sam | 15 | male |
| alex | 29 | male |

Table1: the typical user profile wide table

### 2.2. task manager

The practical industry profile system is quite a complicated one, it involves several specific types of offline tasks, if they just arranged by hand, it exhausted much human effort. All in all, we present a task manager to manage all the tasks fine-grained, such as when scheduling the task, which task needs to run with priority after parent dependencies are completed.

**Task Checker:** Retrieve metadata from meta manger, decide which table is available by checking in fixed cycle, then create a task instance for scheduler to dispatch, and trace the metrics of each task instance and report to **metrics center**.

**Task Scheduler:** Schedule task instance in a fixed period, and each instance represents specified task type (id mapping or build bitmap). Scheduler has the ability to dispatch spark tasks to heterogeneous big data platforms, and extend to new platforms easily.

| profile | value | user bitmap |
| --- | --- | --- |
| age | 15 | [sam] |
| age | 19 | [alex] |
| gender | male | [sam, alex] |

Table 2. the high level bitmap illustration

### 2.3. build engine

The build engine responsible for assembling the core structure of profile label data. Firstly, the id mapping builder used to map string id to number id. Furthermore it is used by the bitmap builder to convert the origin table to distributed bitmap, also it will be described in Section 3. The batch writer interacts with the OLAP store, and writes the distributed bitmap to them.

**Id Mapping Builder:** the user profile data always came from different heterogeneous sources, so the id format might be various. The user id with string type needs to be cast to number for bitmap structure. There is also a challenge that needs to be considered, how to maintain the id be identical for different days mapping, or called consecutive id generator. If it is not the same, the following reader component needs to be modified every day, it's not maintainable and costs much effort. We introduce a novel task to achieve this. The details will be presented in Section 3.

**Bitmap Builder:** the core user profile is typically stored in a raw table like Table 1. Obviously, this method is quite easy to understand, but there are several drawbacks when the scenario comes to a large scale. Firstly, your profile label became more and more. They came from different tables, you need to join tables by the same user key, then get all profiles, this produces additional compute overhead, our hands-on experience demonstrated that it could consume 50% computation resources in your hand when at scale. Secondly, the space might be inefficient rather than bitmap structure, which is shown in Table 2. Therefore, the bitmap structure can compress id in just 1 bit [9].

**Batch Writer:** The unified writer implementation to sink bitmap structure to OLAP engine such as Clickhouse [10]**,** PostgreSQL [11]**,** StarRocks [12]**,** we introduce unified writer layer to hide the difference between each databases, a unified writer api for upper layer, thus the high level build engine didn't care about the real database.

## 2.4. distributed bitmap

The bitmap is our core data structure to store practical industry data sets. Most modern OLAP databases support this kind of structure, and always give better performance than a typical wide table. As mentioned before, the bitmap can avoid join rather than the typical solution, instead we use **OR**, **XOR**, **AND** operations in memory to replace inefficient shuffle communication overhead. There also several challenges we need to consider:

**bitmap across machines**: Real-industry data is really at a large scale, so make data distributed always be a preferred solution, the question is how to distribute the compact bitmap structure, in our solution present a novel concept **tablet**, it is a child table, contains each profile label, and with partial user, as a result, we can distributed our bitmap by user across several machines. Details would be illustrated in Section 3.

## 2.5. Online Service

Whenever profile labels are stored efficiently, or set to the right place, the online service such as analysis tool, indicator query will retrieve data from distributed bitmap, there is also some optimized solution like parallel reading to retrieve as quickly as possible, without joining process the ad hoc query might be more efficient.

## 2.6. metrics center

One principle when we designed our system is that it can be monitored at every corner, from producing to consuming, collecting the table metrics, label metrics and task metrics to show the state in each step from raw data to business value, then we have a dashboard to know the current state of the profile system. Even more, it could alarm immediately when there is a problem, and recover as fast as possible.

# 3. Implementation

The **BitUP** is quite efficient for retrieving label data, compared to previous user profile systems, which mostly use widetable as store layer, it avoids the join process and reduces compute overhead, using distributed bitmap (Section 3.2) to accelerate the join process by **OR, AND, XOR** operations. By introducing the novel concept tablet, this system can parallelize the reading process (Section 3.3) and improve performance greatly. Let's discuss it in detail.

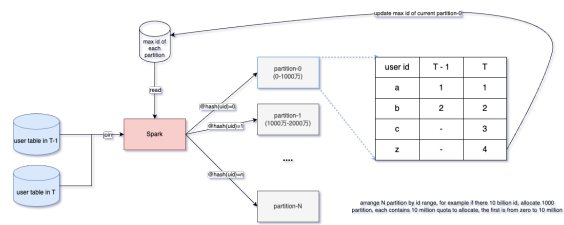

Figure 2: how the consecutive id generator work

### 3.1. consecutive id generator

The first problem we met was how to convert the different id to number so it can fit into a bitmap.

Firstly, we will arrange how many partitions should be allocated by current ranges, for example if there are 10 billion ids, we probably allocate one thousand partitions for each segment of ids, next we'll hash the string id into different partitions. To make it identical to the previous allocated number, we use a join mechanism, combine the two days user table which records the origin id and allocated number id, so if it already has a number id, just keep it. If not, request the id allocator to get a max id in the current partition.

The allocator will increase max id from allocation process, handle the request to get a max id in specified partition.

With the above id generate mechanism, **BitUP** can solve the problem to generate consecutive number id for following bitmap builder. The whole visualize data flow shown in Figure 2.

**Algorithm 1** bitmap builder
1: **for** label = age,gender,... **do**
2:    **for** partition = p0,p1,... **do**
3:       Group user by current label, yield $uid\_array <- u1, u2, ...$
4:       filter $uid$ in $uid\_array$ which must in current partition use $hash(uid) == partition$
5:       Initialize two bitmap named $bits\_1, bits\_2$
6:       **for** uid = u1,u2,... **do**
7:          **if** $uid \wedge 0xffffffff0000000 == 0$ **then**
8:             set the bit of uid in $bits\_1$ to 1
9:          **else**
10:            set the bit of $uid \wedge 0xffffffff0000000$ in $bits\_2$ to 1
11:          **end if**
12:       **end for**
13:       sink two bitmap to tablet with same $partition$
14:    **end for**
15: **end for**

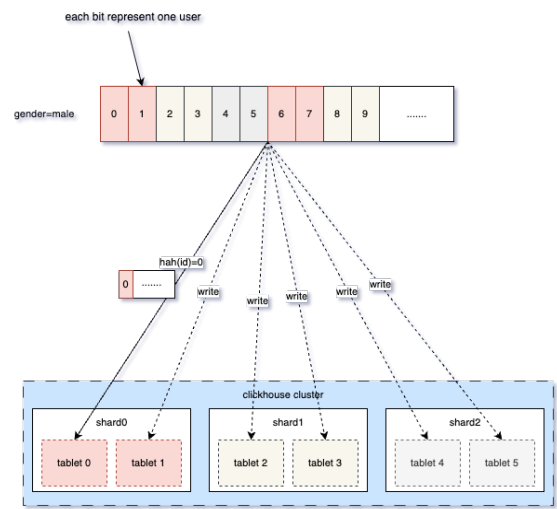

Figure 3. distributed bitmap sink to tablet in clickhouse

### 3.2. distributed bitmap

Next, how to build the core structure distributed bitmap. Firstly in real industry, the profile data is quite large, the user size might exceed billions, so 32-bit range could not be enough, if we just simple use 64-bit solution the space overhead is not acceptable, as we test the performance and space cost is so large when using 64-bit, and also can not be retrieve in seconds, so our plan using two bitmap to split the user into two parts, one for zero to 4 billion ($2^{32}$), the second for 4 billion to 8 billion, when user query profile label, it need to fetch two parts of bitmap and merge to final result.

In addition there is another problem: when the cardinality of bitmap increases significantly, the query performance decreases dramatically, because most current OLAP engines do not parallelize the bitmap operation very well, therefore we need to fine-tune it by ourselves. Moreover in industry dataset our cardinality may exceed billions in just one label such as age, gender, because these common user attributes acquired by a huge user base.

On top of that, we introduce a novel concept to represent each bitmap segment in the OLAP table, we call it tablet, as shown in Figure 3. We hash different users into each bitmap segment, and sink to the tablet in each shard of clickhouse, every tablet stored as a table in clickhouse. The overall algorithm is also illustrated in Algorithm 1.

Using the above method, we solve the bitmap cardinality problem and increase query performance significantly. This evaluation will show in Section 4. Moreover it can scale up linearly for practical datasets at any scale.

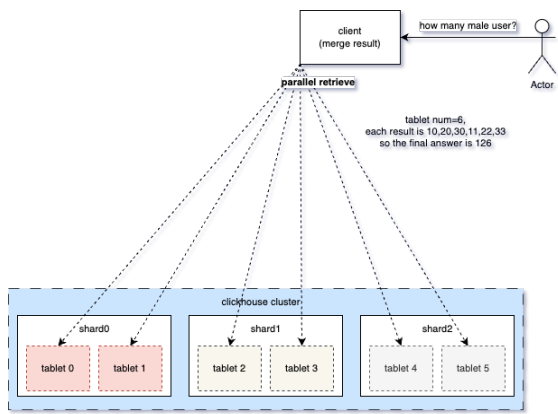

Figure 4. parallel reader for tablet

### 3.3. parallel bitmap reader

After completing the building process of distributed bitmap, the query process needs a design to match the described data structure. We introduce a module named parallel bitmap reader, typically we can answer the query like **how many male users in our business**? From previous build process, the relation data is already stored in a distributed bitmap, so it can be solved by several steps:

- **step1:** retrieve metadata about how many tablet it contains
- **step2:** assume tablet num is 6, parallel send query to 6 tablet, ask the same question **how many male users are there?**
- **step3:** get partial result like 10, 20, 30, 11, 22, 33, then combine it, the final answer is 126

This system can also support other types with multi conditions such as gender=male and age=10. This system is also versatile for download, origin user detail could be downloaded by reverse the mapping strategy, which is stored in our clickhouse cluster. Next we'll discuss our evaluation.

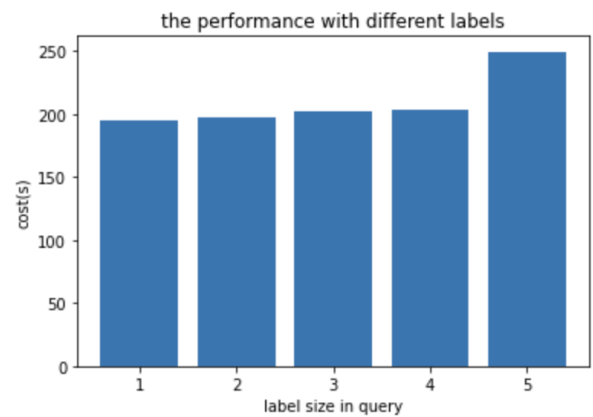

Figure 5. query performance with different label size

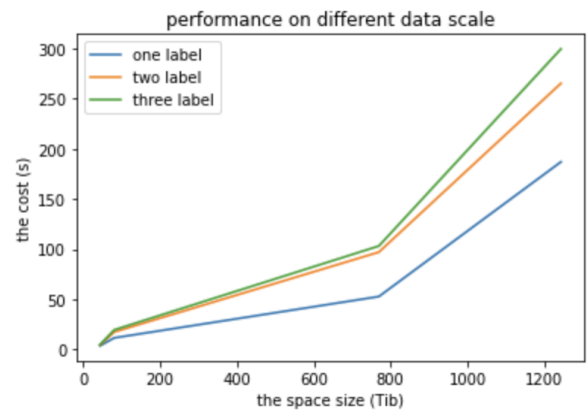

Figure 6. query performance in different data scale

## 4. Experiments

We evaluate our **BitUP** system with two experiments. First with the same data set use a different query to test its performance for scale on label. Second with the same condition running on a data set at a different scale to test its performance for various data scales. We apply our experiments on a clickhouse cluster with 13 worker nodes, each worker has 64 core cpu, 256 Gib memory and SSD.

### 4.1 performance on label size

We design the experiment with different query labels on the same data set to verify the computation overhead on bitmap **AND** operation. With different size of label in query condition, the consumed time didn't increase a lot, thanks to the architecture of distributed bitmap, we can parallelize our query to all tablet which make the bitmap operation pretty fast in each bitmap segment rather than only one huge bitmap, and the optimized index created an tablet also help for locating the file block as fast as possible. The distributed bitmap structure solves the bottleneck of disk io and memory computation, as a consequence in Figure 5, the time cost is quite even across different label sizes.

### 4.2 performance on data scale

To test the scale ability in a large data set, we designed the second experiment. We prepared several data sets from 43 Tib to 1.2 PB, the max dataset has 1.3 trillion rows, which is quite at industry scale. We monitor the cost time in each label query, and the results are shown in Figure 6. Notably, the cost increases quite linearly, from 1 label to 3 label, the performance is quite similar, it proved that distributed bitmaps can scale up very well in quite large datasets.

## 5. Summary

We described a user profile data system using a novel distributed bitmap to accelerate the performance for user query rather than the previous way, breaking the limit for profile store in traditional ways such as widetable. This system is efficient; You can query a label in less than a second. Moreover it provides elastic scalability, in our industry scene could support over 1PB dataset, containing more than ten trillion rows. Also we see the cost increase linearly, the total computation overhead is quite acceptable. Finally, we prepare two experiments to prove our thoughts, the distributed bitmap is as fast as possible.

In our practical industry circumstance, this new generation of user profile data system works pretty well, about thousands people use it everyday, average cost is seconds.